\documentclass[aps,PRL,twocolumn,superscriptaddress,floats]{revtex4}
\usepackage{txfonts}
\usepackage{amssymb}
\usepackage{graphicx}
\usepackage{sidecap}

\begin{document}

\title{An unified minimum effective model of magnetism  in iron-based superconductors}
\author{Jiangping Hu}\email{jphu@iphy.ac.cn/hu4@purdue.edu} \affiliation{Beijing National Laboratory for Condensed Matter Physics and Institute of
Physics, Chinese Academy of Sciences, P. O. Box 603, Beijing 100190, China}
\affiliation{Department of Physics, Purdue University, West Lafayette, Indiana 47907, USA}
 \author{Bao Xu}
  \affiliation{Beijing National Laboratory for Condensed Matter Physics and Institute of
Physics, Chinese Academy of Sciences, P. O. Box 603, Beijing 100190, China}
\author{Wuming Liu}
\affiliation{Beijing National Laboratory for Condensed Matter Physics and Institute of
Physics, Chinese Academy of Sciences, P. O. Box 603, Beijing 100190, China}
\author{Ning-Ning Hao}
   \affiliation{Beijing National Laboratory for Condensed Matter Physics and Institute of
Physics, Chinese Academy of Sciences, P. O. Box 603, Beijing 100190, China}  \author{Yupeng Wang}
  \affiliation{Beijing National Laboratory for Condensed Matter Physics and Institute of
Physics, Chinese Academy of Sciences, P. O. Box 603, Beijing 100190, China}

\maketitle

\textbf{Since 2008, many new families of iron-based high temperature (high-$T_c$) superconductors have been discovered \cite{Hosono,ChenXH,FeTe,ChenXL}. Unlike all parent compounds of cuprates that share a common antiferromagnetically (AF) ordered ground state, those of iron-based superconductors exhibit  many different AF ordered ground states, including collinear-AF (CAF) state in  ferropnictides \cite{caf}, bicollinear-AF (BCAF) state in 11-ferrochalcogenide $FeTe$ \cite{bcaf,bcaf2}, and block-AF (BAF) state in 122-ferrochalcogenide $K_{0.8}Fe_{1.6}Se_2$ \cite{baf}. While the universal presence of antiferromagnetism suggests that superconductivity is strongly interrelated with magnetism, the diversity of the AF ordered states obscures their interplay. Here we show that all magnetic phases can be unified within an effective magnetic model. This model captures three incommensurate magnetic phases, two of which have been observed experimentally. The model characterizes the nature of phase transitions between the different magnetic phases and explains  a variety of magnetic properties, such as  spin-wave spectra and electronic nematism.  Most importantly, by unifying the understanding of magnetism, we  cast new insight on the key ingredients of magnetic interactions which are critical to the occurrence of superconductivity.
}
\\

The iron-based superconductors can be divided  into two major  classes,  ferropnictides and ferrochalcogenides.  Superconductivity  in both classes, like  cuprates,  arises from electron- or hole-doping of their AF parent compounds.   Hence,  determining magnetic interactions  in the parent compounds is extremely important in the identification of   key magnetic interactions relevant to  the development of superconductivity.

However, because of the diversity of the magnetic orders and the fact that the magnetic properties  exhibit the dichotomic behavior of both local moment and itinerant electron aspects,
it has been extremely difficult to  find a consistent magnetic model to describe the magnetism of  iron-based superconductors. Theoretically, magnetism can be explained by either local moment models where local spins interact with each other  or  itinerant electron models where magnetic order arises from  nested Fermi surfaces. The former is appropriate for   insulating materials such as the parent compounds of cuprates  while  the latter is   suitable for metallic systems, such as chromium.  However, the iron-based superconductors include rather diversified materials whose parent compounds can have  either metallic or insulating ground states. More specifically,  the parent compounds of ferropnictides  are bad metal \cite{Hosono} while the newly discovered 122-ferrochalcogenide, $K_{0.8}Fe_{1.6}Se_2$ is a block AF insulator \cite{baf}. Moreover, it was  shown that  even in the insulating parent compounds, itinerant electron aspects have to be included \cite{my} and   in the metallic parent compounds, local moment aspects are manifested \cite{zhao1}.  Such a dichotomy of the magnetic  properties \cite{my,zhao1,it2,it1, lip2, har, dong} leads to many diversified viewpoints  on what is the proper model to describe the magnetism \cite{local1, local2, local3, local4,  it4,it6}.

In ferropnictides, the CAF state as shown in fig1(b), in principle,  can be explained by Fermi surface nesting
between hole pockets at $\Gamma$ and electron pockets at $X$ \cite{dongj} because they are  connected by the CAF ordered wavevector $Q_{CAF}=(0,\pi)$ in unfolded reciprocal space.  However, the mechanism fails to explain the BCAF and BAF states shown in fig1(c,d).  In these two cases, no matching Fermi pockets  can be connected with their ordered wavevectors $Q_{BCAF}=(\pi/2,\pi/2)$
and $Q_{BAF}=(3\pi/5,\pi/5)$.  Within a local moment picture, the CAF state can be naturally obtained in a Heisenberg model with nearest neighbor (NN) $J_1$ and next nearest neighbor (NNN) $J_2$  magnetic exchange interactions if $J_2>|J_1|/2$ \cite{local1,local2, local3}. However, this model is quantitatively  incompatible with the large anisotropy of the NN exchange interactions along the ferromagnetic (FM) and AF directions in the CAF state  determined in neutron scattering experiments \cite{zhao1}. Intriguingly, recent neutron experiments in ferrochalcogenides show that the similar large NN anisotropy exists in both BAF\cite{lip2} and BCAF states \cite{my}. Moreover, there is almost no anisotropy between the NNN exchange interactions even though the magnetic configurations along the two NNN coupling directions in both states are different \cite{lip2,my} as shown in fig1(c,d).  Two completely different solutions backed by electronic structure calculations have been proposed to solve the incompatibility in ferropnictides:  one emphasizes orbital order \cite{lv,turner, leec} and the other suggests  biquadratic spin interaction terms \cite{local4}.  However, both solutions are inadequate in understanding the BAF and BCAF phases. For example, electronic structure calculations and orbital ordering mechanism  also suggested an large NNN anisotropy in ferropnictides \cite{Han, cao}, which is inconsistent with experimental results \cite{my,lip2}.

Here we attempt to formulate a minimum effective  spin model that unites the description of the magnetic properties of  the parent compounds of the different classes of  iron-based superconductors. The model has to preserve the tetragonal lattice symmetry so that  it is capable of providing us the detailed relations between different magnetically ordered  states as consequences of spontaneous symmetry breaking at low temperature.  The model should be able to capture   all the magnetically ordered ground states observed in iron-based superconductors, to explain their correct spin-wave spectra and  the anisotropy of magnetic exchange interactions, and to predict possible new states including incommensurate magnetic states.   In the following, we will show by including the NN biquadratic interaction term  as proposed in ref.\cite{local4}, but not the NNN biquadratic interaction term,  and the next next nearest neighbor (NNNN) AF Heisenberg interactions $J_3$ \cite{ma, cfang}  in $J_1-J_2-J_c$ model \cite{local2},  we can fulfill above requirements.\\
\\
{\bf Model Hamiltonian}
 We start with the following general Hamiltonian,
  \begin{eqnarray}
    H=\sum_{ij,n}[J_{ij}\vec{S}_{i,n}\cdot\vec{S}_{j,n}-K_{ij}(\vec{S}_{i,n}\cdot\vec{S}_{j,n})^2]
                 +J_c\sum_{i,n}\vec{S}_{i,n}\cdot\vec{S}_{i,n+1}
  \end{eqnarray}
where $J_{ij}$ describe in-plane magnetic exchange interactions, $J_c$ is  inter-plane magnetic coupling along c-axis (between iron layers) and $K_{ij}$ are  in-plane non-Heisenberg biquadratic  couplings. In  the minimum model proposed here, we choose
nonvanishing $J_{ij}$ = $J_1$,  $J_2$  or  $J_3$  if and only if $i,j$  are  two NN, NNN, or  NNNN  sites respectively and   $K_{ij} =K$ if and only if $i,j$ are two NN sites.   The interactions  are sketched in fig1(a) by the dashed lines.

We note  that the model is a natural extension of  models in ref.\cite{local2, local4,ma, cfang} proposed for ferropnictides  and  ferrochalcogenides before. However, all previous models only describe particular family and fail to provide a comprehensive understanding of  different magnetic states. \\
 \\
{\bf Exact  Classical Phase Diagram}
 The classical  phase diagram of the model can be obtained exactly.  In fig2,  we draw  a typical phase diagram  in the  $J_3/|J_1|$-$J_2/|J_1|$ plane by taking $KS^2/|J_1|=0.2$.  The phase diagram is almost symmetric between  $J_1>0$ (the right part of fig2) and $J_1<0$ (the left part of fig2). For $J_1>0$, there are three commensurate phases labeled as AFM, CAF and BCAF in fig2,  which exactly describe the static magnetic states of the parent compounds of curpates, ferropnictides and 11-ferrochalcogenide $FeTe$ respectively.  There are also two incommensurate phases sandwiched between the commensurate phases with  ordered incommensurate wavevectors $(q,\pi)$ or $(\pi,q)$ (labeled as IC1) and $(q,q)$ (labeled as IC3) respectively.  The static $(q,q)$ phase was observed in $Fe_{1+y}Te$ when $y>0.1$ \cite{bcaf}.  Although  no static  $(q,\pi) $  phase has been detected,  the $(q,\pi)$ incommensurate spin fluctuations have  been observed in $FeTe_{1-x}Se_x$ \cite{lum, arg} ,  electron-overdoped $Ba(Fe_{1-x}Co_x)_2As_2$ \cite{pratt}, and hole-doped $KFe_2As_2$ \cite{icm1}.
 If $J_1$ is switched to negative, namely ferromagnetic (FM),  the AFM phase becomes a FM phase and $(q,\pi)$ becomes $(q,0)$ or $(0,q)$ (labeled as IC2).

More specifically, we can determine phase transition boundary.  We scale other parameters with $J_1$ as  $\tilde K=KS^2/J_1$, $\tilde{ J}_2=J_2/J_1$,
  and $\tilde{J}_3=J_3/J_1$ for simplicity.  The phase boundary between the BCAF and  the $(q,\pi)$ incommensurate phase is determined by $4(\tilde{J}_3-\frac{1}{4})^2-(\tilde{J}_2-\frac{1}{2})^2=(\tilde{K}-\frac{1}{2})^2$ which defines the upper branch of a hyperbolic curve  centered at $(\tilde{J}_2,\tilde{J}_3)=(1/2,1/4)$.  The
 phase boundary between the AFM and BCAF phases  is determined by  $\tilde{J}_3=-\frac{1}{2}\tilde{J}_2+\frac{1}{2}$ and that betwee
 BCAFM and CAFM phases is determined by   $\tilde{J}_3=\frac{1}{2}\tilde{J}_2$.  The incommensurate states appear only when $\tilde{K}<0.5$. We  emphasize that finite positive $\tilde{K}>(1.5-\sqrt{2})$ and  $\tilde{J}_3>0.25$ are necessary conditions for  the appearance of the BCAF phase.    The incommensurate wavevectors of the three incommensurate phases can also be explicitly determined: (1) $(q,0)$ or $(0,q)$ phase with $q={\rm arccos}\frac{2\tilde{J}_2+1}{2\lambda}$;  (2) $(q,\pi)$ or $(\pi,q)$ with $q={\rm arccos}\frac{1-2\tilde{J}_2}{2\lambda}$;  and
   (3) $(q,q)$ with $q={\rm arccos}\frac{1}{2(\tilde{J}_2-\lambda)}$, where $\lambda=\tilde{K}-2\tilde{J}_3$.  \\
 \\
 {\bf  Commensurate Phases}
 The model captures three commensurate phases: AFM, CAF and BCAF.  In  these commensurate phases, the biquadratic interaction term  effectively creates the anisotropy of the NN magnetic exchange interactions by taking a meanfield decoupling \cite{local4}.  Depending on the spin alignment of  two NN sites,    $J_{1a}=J_1+2KS^2$  if it is AF and   $J_{1b}=J_1-2KS^2$ if it is FM.  Therefore,  effectively,  our model becomes a $J_{1a}-J_{1b}-J_2-J_3-J_c $ model in these phases.  Experimentally, the spin wave excitations in the parent compounds, $Ca Fe_2As_2$ \cite{zhao1} and $BaFe_2As_2$ \cite{har}, were fitted well to the  $J_{1a}-J_{1b}-J_2-J_c$ model. 
 and those of $FeTe$ was fitted well to a $J_{1a}-J_{1b}-J_2-J_3-J_c$ model \cite{lip2}.

 More interestingly, in  the ferrochalcogenide $K_{0.8}Fe_{1.6}Se_2$, featuring intrinsic vacancy ordering,  the fitting of spin wave excitations  to a $J_{1a}-J_{1b}-J_2-J'_2-J_3$ model has concluded a large FM  $J_{1a}$,   AF $J_{1b}$ and  very small anisotropy between two AF  NNN couplings $J_2$ an $J'_2$\cite{my}.   Our model hence   successfully describes the magnetism of this new material as well. The presence of vacancy ordering does not  drastically affect the magnetic interactions.  Instead, it reduces the magnetic frustration to stabilize the BAF order \cite{fangx}.

 From the experimental results of spin wave excitations \cite{zhao1,har,lip2,my},  we now can extract  the magnetic exchange parameters of  our model for different parent compounds. The results are summarized in  Table1. We note that FeTe is near the boundary of the BCAF phase  and incommensurate phases. The values listed in Table1 is within the error bar of experimental values in \cite{lip2}.  This table displays  a central message that  all iron based superconductors share a similar    AF  NNN   exchange interaction $J_2$.  However the sign of $J_1$ is  different  between ferropnictides and ferrochalcogenides  and a significant  AF $J_3$ exists in ferrochalcogenides but not in ferropnictides.   
 \begin{center}
  \begin{table}
  \begin{tabular}{|r|c|c|c|c|c|c|l|}\hline
 Material &  Phases     &  $(Q_x,Q_y)$     &   $ J_1 S   $  & $J_2  S$ & $J_3 S$ & $KS^2 $ & $J_cS$ \\
 \hline
$CaFe_2As_2$ &  CAF       &   $(0,1)\pi$        &  22   & 19 & -- & 14 &5     \\
\hline
$BaFe_2As_2 $&  CAF       &   $(0,1)\pi$        &  25  &   14 & -- & 17 &2    \\
\hline
$FeTe$ & BCAF               & $(\frac{1}{2},\frac{1}{2})\pi$        & $-34$ & $18.5$ & $9.5$ &  9 & --    \\
\hline
$K_{0.8}Fe_{1.6}Se_2$ & BAF    & $(\frac{3}{5},\frac{1}{5})\pi$        & -10 & 16 & 9 &  12 & 1.4    \\
 \hline 
 \end{tabular}
 \label{parameters} 
 \caption{ The values of the magnetic exchange interactions in  $J_1-J_2-J_3-K$  model obtained from the experimental results of  different parent compounds of iron-based superconductors \cite{zhao1, har, lip2, my}. }
 \end{table}
 \end{center}
{\bf Spin Excitations  in Incommensurate Phases}  The model also predicts three incommensurate phases. We can calculate the distinct features of spin excitations in these three phases.  The typical  spin excitation spectra  in the three incommensurate phases as the function of  wavevectors $\delta$  are  shown in fig3(a)-(f).
 In general,  spin excitations include two branches, which can be identified as an acoustic mode and an optical mode.  In the $(\pi, q)$ phase, the two modes connect with each other at two new  incommensurate positions $\delta=\pm q_0$ when the wavevector   $( \pi,\delta)$ varies as shown fig3(a). If we calculate the intensity of these spin wave excitations measured by neutron scattering,   an hour-glass like behavior along this direction becomes prominent at low energy as shown in fig4.   This behavior has been recently reported in $FeSe_{0.4}Te_{0.6}$\cite{sl} and a clear explanation was not given before.  The dispersion along $(\delta,\pi)$ as varying $\delta$  has much larger energy dispersion than the one along $(\pi,\delta)$ and reaches maximum at $(\pi,\pi)$ which is consistent with experimental results observed in ref.\cite{lum} as shown in fig3(b).  In $(0,q)$ phase,  the dispersion of spin excitations along $(0,\delta)$ as varying $\delta$ is  very similar to the one along $(\delta, \pi)$ in $(\pi, q)$ phase as shown in fig3(c) and that along $(\delta,0)$ is similar to the one along $(\pi, \delta)$ as shown in fig3(d).  In the (q,q) phase,  the two spin wave modes connect at $(0, \pi)$ point as shown in fig3(e,f) and fig3(g,h),  displaying  a Dirac-type dispersion. The two modes has much larger energy separation in the case of $J_1>0$ (shown in fig3(e,f) ) than $J_1<0 $ (shown in fig3(g,h).   More detailed spin wave properties are included in supplement materials. These distinct features can be used to determine the effective magnetic exchange couplings even if the incommensurate order is not static.
\\
 \\
{\bf Nematism and Effective Field Theory}
Both the CAF and BCAF states break the $C^4$ rotational symmetry of the tetragonal lattice.  The rotational symmetry breaking can be separately described by an Ising or nematic order as shown in ref.\cite{local2,local3}.  Without the specific biquadratic term $K$,  when the parameters of the $J_1-J_2-J_c$ model are fixed in the CAF phase region, a weak biquadratic term can be developed through the `order by disorder' mechanism \cite{chandra, local2,local3}  and the nematic phase transition can take place at a transition temperature $T_N$ higher than the CAF transition temperature $T_c$  if the inter-layer coupling $J_c$ is much weaker than  $J_2$ \cite{local2} .   This physics can be analytically described in the continuum limit.   As show in ref.\cite{local2},  the effective field theory of the $J_1-J_2-J_c$ model in the continuum limt is given by
\begin{widetext}  \begin{eqnarray}
 H_{CAF}= \int d^2{\bf r}\sum_{n,\alpha}\Big[\frac 1 2   J_2|\nabla {\vec \phi}_{n,\alpha}({\bf r})|^2
- J_c
\vec\phi_{n,\alpha}({\bf r})\cdot \vec \phi_{n+1,\alpha}({\bf r})\Big]
 - g\sum_{n}\left[\vec \phi_{n,1}({\bf r})\cdot {\vec \phi}_{n,2}({\bf r})\right]^2
 +  J_1\sum_{n}{\vec \phi}_{n,1}({\bf
r})\partial_x\partial_y{\vec\phi}_{n,2}({\bf r}),~~~~~~ \end{eqnarray}
\end{widetext}
where  we use the same notions as ref. \cite{local2}:  $\vec \phi_{n,\alpha=1,2}$  specify the two AF Neel orders in the two sublattices of the tetragonal lattice shown in fig1(a) ( for simplicity, we take $S=1$ in this section).  The nematic order is defined to be $\sigma =2g \langle \vec \phi_{n,1}({\bf r})\cdot\vec
\phi_{n,2}({\bf r})\rangle $.  Without the  biquadratic term $K$, $g\sim 0.13  J_1^2/ J_2$.  With this term, we just need to modify $g\sim0.13  J_1^2/ J_2+ K$.
Therefore, the calculations and the results in ref. \cite{local2} are still valid. The large $g$ value due to the specific biquadratic $K$ term  simply enhances the nematic order and increases the temperature range of spin nematic fluctuation  above $T_N$, which has been observed experimentally\cite{har}. 

In the case of BCAF states, the similar effective model for $J_1-J_2-J_3$ model has also been derived in ref.\cite{xuhu} . Including $J_c$, the effective field theory of the $J_1-J_2-J_3-J_c$ model close to the BCAF states can be written as
\begin{widetext}
\begin{eqnarray}
 H_{BCAF} &= &\int d^2{\bf r}\sum_{n,\alpha}\Big[\frac 1 2   J_3|\nabla {\vec \phi}_{n,\alpha}({\bf r})|^2
-  J_c
\vec\phi_{n,\alpha}({\bf r})\cdot \vec \phi_{n+1,\alpha}({\bf r})\Big] - g'\sum_{n}\{\left[\vec \phi_{n,1}({\bf r})\cdot {\vec \phi}_{n,3}({\bf r})\right]^2+\left[\vec \phi_{n,2}({\bf r})\cdot {\vec \phi}_{n,4}({\bf r})\right]^2\}\nonumber \\
& & -  J_1\int d^2{\bf r}\sum_{n}\Big [\vec{\phi}_{n,1} ({\bf r})\cdot \nabla_x \vec{\phi}_{n,2} ({\bf r})+
 \vec{\phi}_{n,4}({\bf r})\cdot \nabla_x \vec{\phi}_{n,3}({\bf r}) -
\vec{\phi}_{n,2}({\bf r}) \cdot \nabla_y \vec{\phi}_{n,3}({\bf r}) - \vec{\phi}_{n,1}({\bf r})
\cdot \nabla_y \vec{\phi}_{n,4} ({\bf r})\Big]
 \end{eqnarray}
 \end{widetext}
where $\vec \phi_{n,\alpha}$ with $\alpha=1,2,3,4$ are four Neel order
parameters defined in the four sublattices of the tretragonal lattice specified by  the $J_3$ coupling  and  $g' \sim 0.13  J_2^2/  J_3$. Differing from the case of  the CAF,  the biquadratic term in the case of  the BCAF  adds an additional coupling term into above effective Hamiltonian, which is given by
\begin{widetext}
\begin{eqnarray}
H_{K}=-  K \int d^2{\bf r}\sum_{n}\{\left[\vec \phi_{n,1}({\bf r})\cdot {\vec \phi}_{n,2}({\bf r})\right]^2+\left[\vec \phi_{n,1}({\bf r})\cdot {\vec \phi}_{n,4}({\bf r})\right]^2+
\left[\vec \phi_{n,2}({\bf r})\cdot {\vec \phi}_{n,3}({\bf r})\right]^2+\left[\vec \phi_{n,3}({\bf r})\cdot {\vec \phi}_{n,4}({\bf r})\right]^2\}\end{eqnarray}
\end{widetext}
The complete field description of the $J_1-J_2-J_3-J_c-K$ model near the $BCAF$ phase is given by $H_{T}= H_{BCAF}+H_{K}$.  In this model, there are four independent Ising orders $\sigma_{1}=g'\vec \phi_{n,1}({\bf r})\cdot {\vec \phi}_{n,3}({\bf r})$, 
$\sigma_{2}=g'\vec \phi_{n,2}({\bf r})\cdot {\vec \phi}_{n,4}({\bf r})$, $\sigma_{+}=K(\vec \phi_{n,1}({\bf r})+  {\vec \phi}_{n,3}({\bf r}))\cdot (\vec \phi_{n,2}({\bf r})+  {\vec \phi}_{n,4}({\bf r}))$ and  $\sigma_{-}=K(\vec \phi_{n,1}({\bf r})-  {\vec \phi}_{n,3}({\bf r}))\cdot (\vec \phi_{n,2}({\bf r})+ {\vec \phi}_{n,4}({\bf r}))$.  A detailed study of this  effective field model will be present elsewhere.  Here, we simply discuss the case  when $K$ is much larger than $g'$ to demonstrate that the effective field model indeed captures phase transitions.  In this case,   the transition between the BCAF state and the (q,q) incommensurate state is controlled by the Ising order $\sigma_{+}$.   It is easy to show that when $|\sigma_+|$ is large, the ground state is the BCAF state. The transition takes place at  $|\sigma_c|=\frac{J_1^2}{4J_3^2}$. When $|\sigma_+|<|\sigma_c|$,  the model is in an incommensurate $(q,q)$ state with $q=\frac{2}{J_1}(|\sigma_c|-|\sigma_+|)$. 
\\ 
\\
{\bf Discussion}
By describing the magnetism of  the different parent compounds of iron based superconductors in a single effective magnetic model,   we  can cast  new insight on the microscopic origin of magnetism. From the magnetic exchange coupling paramters of the effective model,  it is very clear  that the magnetism is neither purely local nor purely itinerant, rather it is a complicated mix of the two.  The presence of significant NNNN coupling $J_3$  suggests the local exchange mechanisms such as superexchange or double exchange
are not enough to account for all magnetic exchange interactions.  Moreover, the sign change of $J_1$ between ferropnictides and ferrochacogenides suggests that the NN exchange interactions are sensitive to subtle difference in band structures.  However,  the robustness of NNN $J_2$ interactions indicates that the NNN $ J_2$ coupling is most likely determined by local superexchange mechanism.   In ref.\cite{local4}, the authors suggest that   the NN biquadratic term $K$ in ferropnictides stems from the strong magnetostructural coupling. However,  this  explanation does not provide an understanding of the absence of NNN biquadratic term in  the BCAF state of $FeTe$  since the lattice distortion in $FeTe$ is monoclinic rather than orthorhombic as in ferropnictides\cite{bcaf}.  More throughout studies of the origin of the biquadratic term 
are needed.

The model reveals the significant difference between ferropnictides and ferrochacogenides:  the sign difference of $J_1$ and the large AF $J_3$ in ferrochcogenides.  These significant differences may suggest the importance of the $p$ orbitals of As or $Te/Se$  on the influence of magnetism. So far, most theoretical models are constructed based on the $d$ orbitals of irons with onsite interactions.  Since    the effect of electron-electron correlations is  believed to be weaker in  ferropnictides than in ferrochalcogenides, one would expect the range of magnetic interactions  should be shorter in ferrochalcogenides than in ferropnictides, which contradicts   the existence of large $J_3$ interactions in ferrochalcogenides but not in ferropnictides.   This contradictory can be resolved  if the significant parts of magnetic exchange interactions are generated through the p-orbitals of As or Se/Te.  The effective  magnetic exchange interactions obtained from onsite electron-electron interactions are not enough to account for   entire magnetic exchange couplings. This also explains that why magnetism is so sensitive to the distance of As or Se/Te away from iron planes \cite{kuroki} because the distance may strongly affect the mixture of $p$ orbitals in electronic structure. 

The model, as an effective low energy model of magnetism,  also tells us the power and limitation of LDA calculations performed for iron-based superconductors where  the effect of electron-electron correlation can not be ignored.
Without any doubt, the LDA calculations explain many magnetic properties in iron-based superconductors. For ferropnictides,  the LDA results  of $J_{1a}$, $J_{1b}$  and $J_2$ values  are in a good agreement with experiments \cite{it6}.
However, LDA calculation wrongly predicted the large anisotropy of $J_2$ in the BCAF and BAF states \cite{Han,cao}. This failure is not surprising  since   the LDA  in magnetically ordered state  is simply  a complicated meanfield approach. 

It has been shown that the high energy magnetic excitations in electron-doped ferropnictides are very similar to those of parent compounds\cite{pcdai}. This proves that the short range magnetic correlations in superconducting states are still dominated by the magnetic exchange interactions determined in the corresponding parent compounds. The doping destroys the long range magnetic correlation but not the short range interactions. Especially, the $J_2$ magnetic exchange interactions should be expected to vary little against doping.  In $FeTe_{1-x}Se_x$,   the incommensurate spin excitations are rather robust against the replacement of Te by Se\cite{lum,arg}. This fact suggests that $J_3$ is relatively stable against the replacement in this family of materials as well.  Therefore, if AF exchange couplings are responsible for superconductivity, we expect both $J_2$ and $J_3$ plays a significant role in superconductivity of ferrochacolgenides.

 \begin{figure*}[]
     \includegraphics[ width=14cm ]{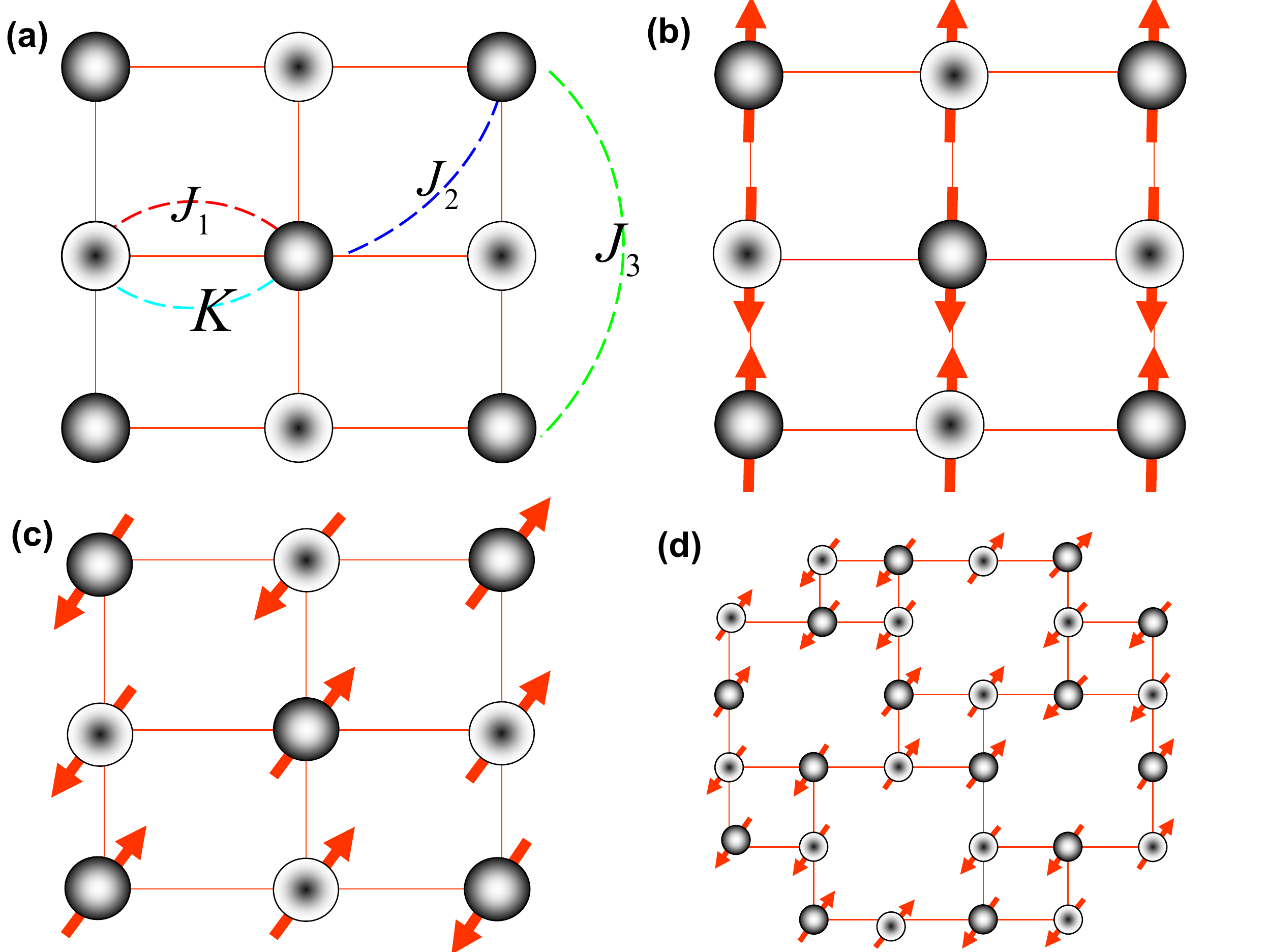}
        \caption{(a). The sketch of magnetic interaction parameters of $J_1-J_2-J_3-K$. (b) The collinear-antiferromagnetic state (CAF) in ironpnictides, for example $CaFe_2 As_2$. (c) The bicollinear-antiferromagnetic state (BCAF) in $FeTe$. (d) The block-antiferromagnetic state (BAF) with $\sqrt{5}\times\sqrt{5}$ vacancy ordering in $K_{0.8}Fe_{1.6}Se_2$. }\label{fig22}
   \end{figure*}

   \begin{figure*}
     \includegraphics[ scale=0.5 ]{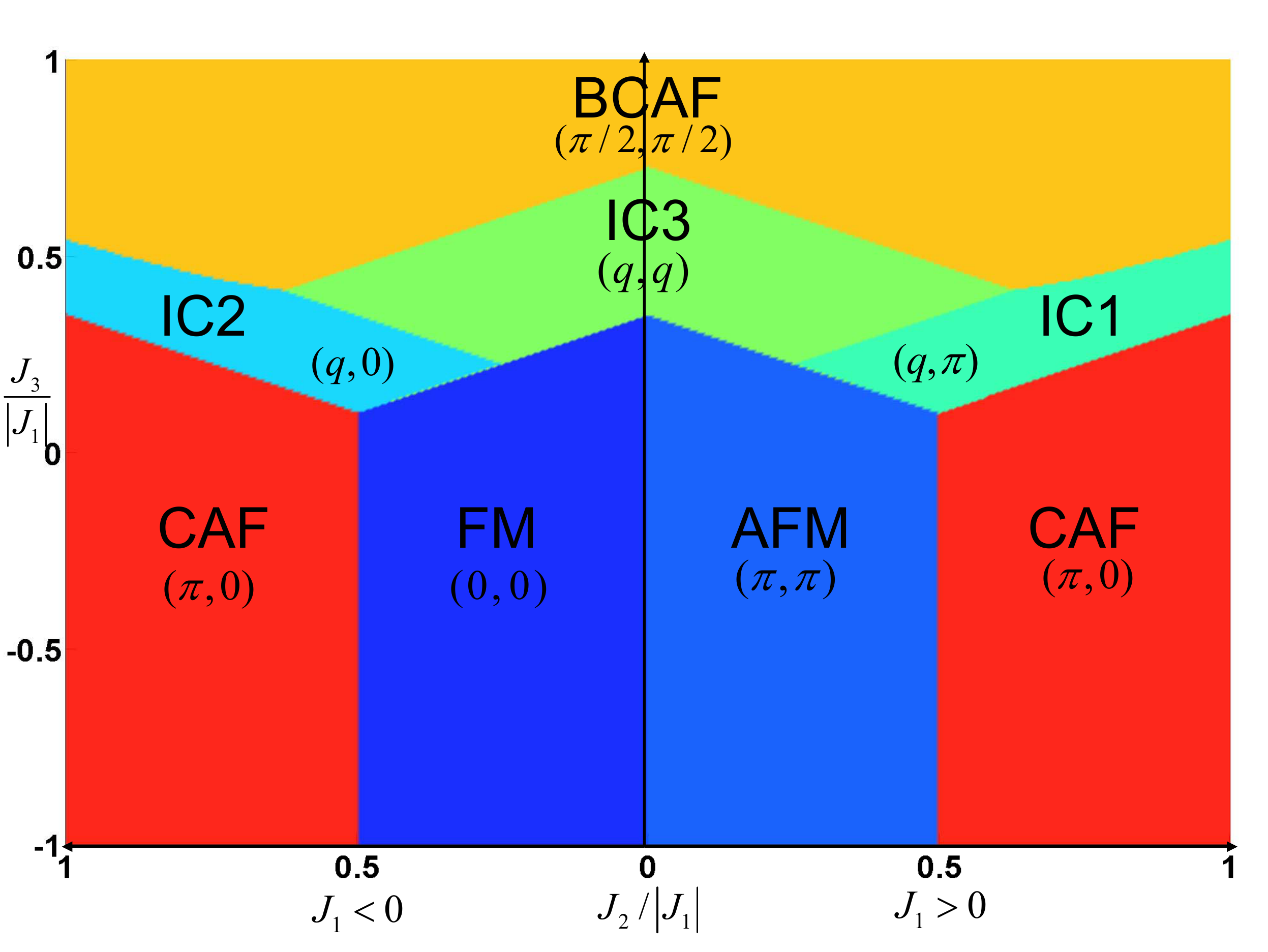}
        \caption{The classical phase diagram of  the $J_1-J_2-J_3-K$ model at $ K=0.2J_1$.  There are total four commensurate and three incommensurate magnetic phases. Their labels and the ordered wavevectors are specified in the figure.
          }\label{fig21}
   \end{figure*}

   \begin{figure*}
     \includegraphics[ scale=0.6]{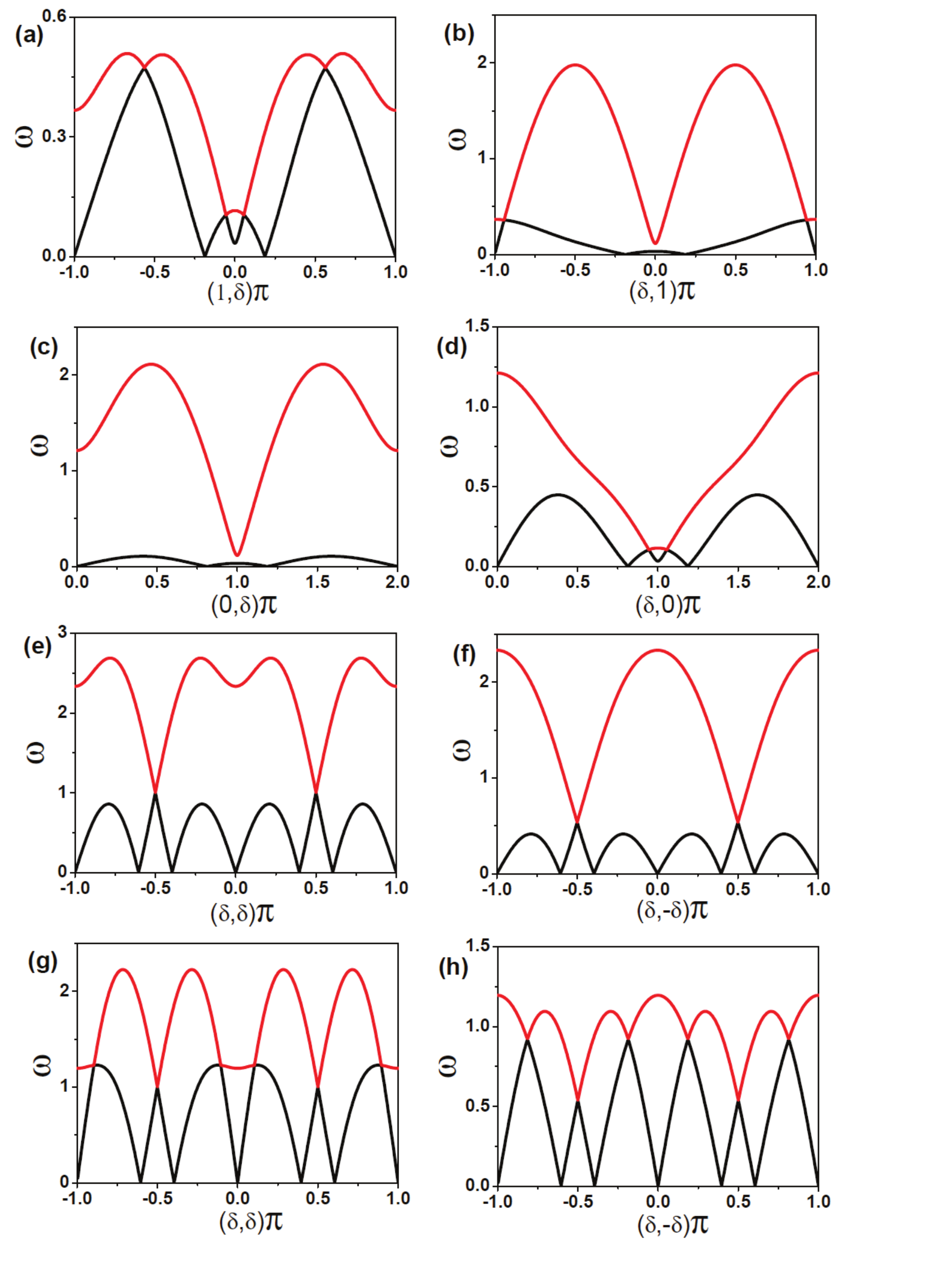}
          \caption{ The spin waves in incommensurate phases:
        (a) along $(1,\delta)\pi$ and (b) along $(\delta,1)\pi$ in the $(\pi,q)$ phase,
        (c) along $(0,\delta)\pi$ and (d) along $(\delta,0)\pi$ in the $(0,q)$ phase,
        and (e)/(g) along $(\delta,\delta)\pi$ and (f)/(h) along $(\delta,-\delta)\pi$ in the $(q,q)$-phase.
        The parameters are fitted to be $S=1$,  $(J_1,J_2/|J_1|,J_3/|J_1|,KS^2/|J_1|)=(1,0.6,0.06,0)$ in (a) and (b),
        $(-1,0.6,0.06,0)$ in (c) and (d), $(-1,0.6,0.5,0.05)$ in (e) and (f), and $(1,0.6,0.5,0.05)$ in (g) and (h).
        }\label{sw}
   \end{figure*}

 \begin{figure*}
     \includegraphics[ scale=0.5 ]{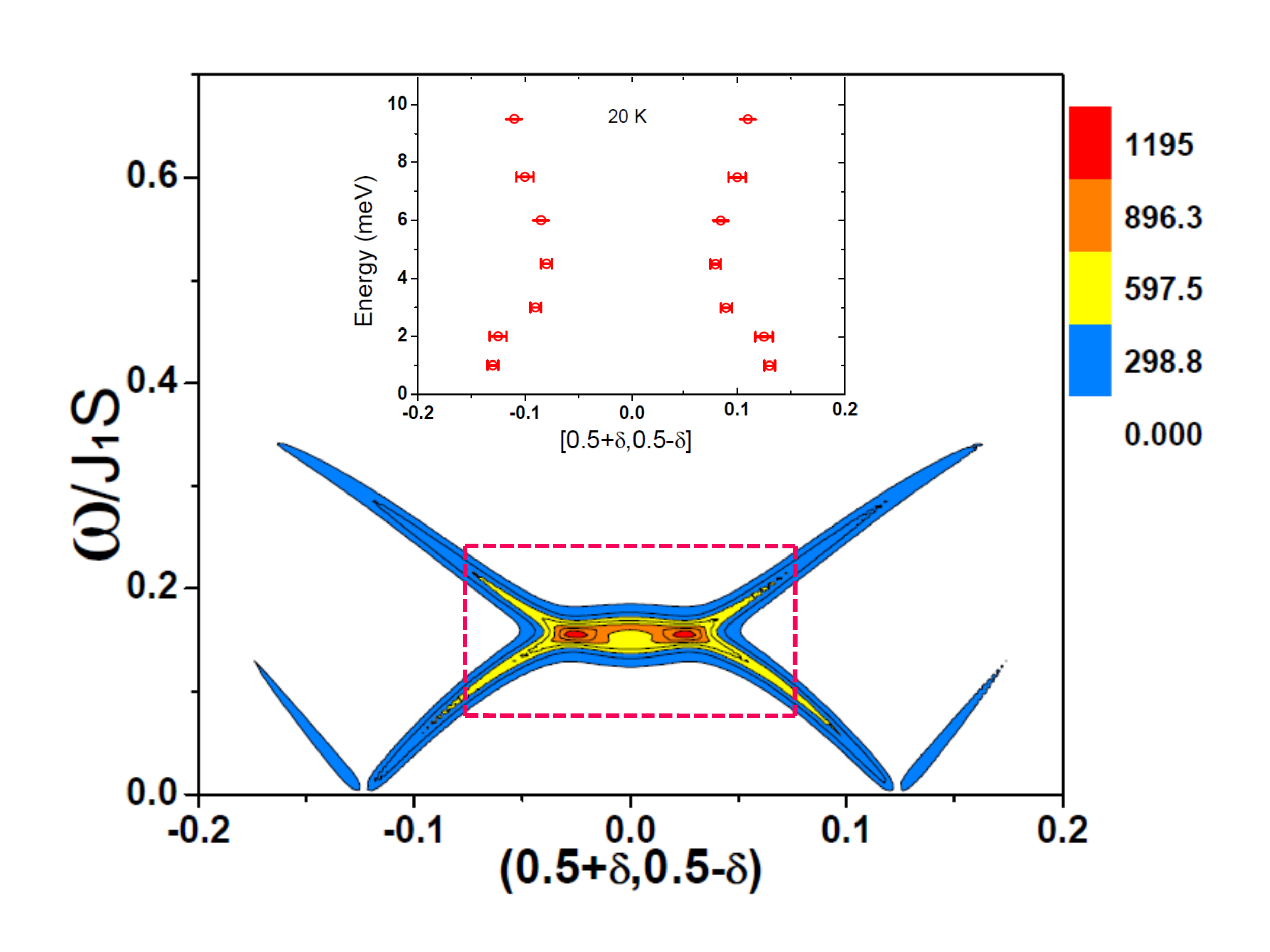}
       \caption{ The hour-glass-like spin waves along $(1/2+\delta,1/2-\delta)_{\rm T}$ in the $(\pi,q)$ phase. 
The parameters are set to $(J_1S,J_2/J_1,J_3/J_1,KS^2/J_1)=(30,0.55,0.045,0.02)$.        
The inset depicts the experimental results released in Ref.\cite{sl}.  
The observed region is labeled by the dashed lines. }
   \end{figure*}

\textbf{Author contributions}
J.P.H. initiates the project and the model described in this paper. J.P. H., B.X.  and N.N.H. carry out  the calculations in this paper.  All of the authors discuss the project and are involved in writing the paper.\\

\textbf{Acknowledgement: }We are grateful to Prof. S. Kivelson,  Prof. Pengcheng Dai,  Prof. Donglei Feng, Prof. Hong Ding, Prof. X. H. Chen, Prof. Tao Xiang and  Prof. Nanlin Wang for fruitful discussions.
This work is supported in part by the National Science Foundation of China  and National Basic Research Program of China (973 Program).

\end{document}